\documentclass[aps,pre,twocolumn,superscriptaddress,showpacs]{revtex4-2}
\usepackage{graphicx}
\usepackage{array}
\usepackage{amsfonts}
\usepackage{amssymb,amsmath,multirow,rotate}
\usepackage{mathrsfs}
\usepackage{booktabs}
\usepackage{threeparttable}
\usepackage{multirow}
\usepackage{epsfig}
\usepackage{threeparttable}
\usepackage{chngpage}
\usepackage{latexsym}
\usepackage{dcolumn}
\usepackage{graphics,graphicx,bm,fleqn,epic,eepic,float}
\usepackage{verbatim}
\usepackage[dvipsnames]{xcolor}
\usepackage{xr}
\usepackage{float}
\usepackage{listings} 
\usepackage{placeins} 
\usepackage{tikz}
\usepackage{amsmath, nccmath}
\usepackage{soul}
\usepackage{blindtext}

\definecolor{red}{rgb}{1,0,0}

\definecolor{blue}{rgb}{0,0,1}

\begin{document}
\setstcolor{red}

\title{Disorder-mediated synchronization resonance in coupled semiconductor lasers} 

\author{Li-Li Ye}
\affiliation{School of Electrical, Computer and Energy Engineering, Arizona State University, Tempe, Arizona 85287, USA}

\author{Nathan Vigne}
\affiliation{Department of Applied Physics, Yale University, New Haven, Connecticut 06520, USA}

\author{Fan-Yi Lin}
\affiliation{Department of Applied Physics, Yale University, New Haven, Connecticut 06520, USA}
\affiliation{Institute of Photonics Technologies, Department of Electrical Engineering, National Tsing Hua University, Hsinchu 30013, Taiwan}

\author{Hui Cao} 
\affiliation{Department of Applied Physics, Yale University, New Haven, Connecticut 06520, USA}

\author{Ying-Cheng Lai} \email{Ying-Cheng.Lai@asu.edu}
\affiliation{School of Electrical, Computer and Energy Engineering, Arizona State University, Tempe, Arizona 85287, USA}
\affiliation{Department of Physics, Arizona State University, Tempe, Arizona 85287, USA}

\date{\today}

\begin{abstract}

	Disorder can profoundly influence synchronization in networks of nonlinear oscillators, sometimes enhancing coherence through external tuning. In semiconductor lasers, however, achieving high‑quality {\rm steady‑state} synchronization is desired, while intrinsic and typically uncontrollable disorder poses a major challenge. Under fixed frequency disorder, we investigate homogeneous fully coupled external‑cavity semiconductor lasers governed by the complex, time‑delayed Lang-Kobayashi equations with experimentally relevant parameters and identify an optimal coupling strength that maximizes steady-state synchronization in the weak‑coupling regime, which we term disorder-mediated synchronization resonance. This optimum appears for any fixed configuration of intrinsic frequency detuning and scales inversely with the number of lasers, leading to a linear scaling of the total coupling cost with the number of lasers. A theory based on an effective thermodynamic potential explains this disorder-mediated optimization, revealing a general mechanism by which moderate coupling can overcome static heterogeneity in nonlinear physical systems.

\end{abstract}

\maketitle

\section{Introduction} \label{sec:intro}

Synchronization in complex dynamical systems has been a topic of continuous
interest~\cite{Kuramoto:book,PRK:book}. A common setting is coupled dynamical
oscillators, where the bifurcation parameter is the coupling strength among 
the oscillators. A focus of many previous studies was on identifying the critical 
point at which a transition from desynchronization to synchronization occurs. 
Depending on the dynamics of the oscillators and the coupling function, the system 
can have a sequence of transitions, giving rise to characteristically distinct 
synchronization behaviors in the parameter space. For example, for a system of 
coupled identical nonlinear oscillators, complete synchronization can arise when 
the coupling exceeds a critical strength as determined by the master stability 
function~\cite{PC:1998,HCLP:2009}. 
In networks of identical chaotic oscillators (without spatial disorder), synchronization can be made most stable by appropriately tuning the coupling strength~\cite{pecora1997fundamentals}, an effect reminiscent of a resonance and well described by the master-stability-function framework. Our problem, however, is fundamentally different: we aim to optimize steady-state synchronization in coupled semiconductor lasers with strong intrinsic disorder, a regime in which the master-stability theory no longer applies.
Systems of phase oscillators with nonlinear
coupling, e.g., those described by the classic Kuramoto model~\cite{Kuramoto:book}, 
can host phase synchronization and the critical coupling strength required for the 
onset of this type of ``weak'' synchronization can be determined by the mean-field 
theory~\cite{WS:1993,OA:2008}. 
A recent study~\cite{mikaberidze2025emergent} employed a gradient‑based optimization approach to identify an optimal sparse coupling structure that maximizes synchrony in the disordered Kuramoto model.
Synchronization in coupled oscillators was 
experimentally studied~\cite{WMRSDS:2013,WSMR:2013}. A counterintuitive 
phenomenon is that adding connections can hinder network synchronization of 
time-delayed oscillators~\cite{HPPMR:2015}. Biomedical applications of 
synchronization have also been actively studied~\cite{JUOLFCF:2017}.

There is a large body of literature on synchronization in networks of nonlinear oscillators~\cite{tang2014synchronization,boccaletti2018synchronization} (see Sec.~\ref{sec:syn_SR} for more background materials). A remarkable phenomenon is that disorders, e.g., random parameter heterogeneity among the oscillators, can counter-intuitively maximize~\cite{BDW:2006} and promote~\cite{ZM:2017,ZOKM:2017} synchronization. For coupled chaotic oscillators, parameter regimes can arise where the oscillator heterogeneity leads to synchronization for conditions under which identical oscillators cannot be synchronized~\cite{SZM:2021}. Synchronization in networked systems with large parameter mismatches was studied in terms of the stability of the synchronous state and transition~\cite{NPS:2023}. Of physical significance with considerable theoretical and experimental interests are external-cavity semiconductor lasers~\cite{LK:1980,winful:1990,lenstra:1991,alsing:1996,DLGK:2000a,DLGK:2000b,DLGK:2001,PLGK:2001,PLGK:2003a,PLGK:2003b,kim:2014,argyris:2016,NHBWB:2021,zhang:2022isochronous,tiana:2022,niiyama:2022,spitz:2023,koyu:2023,NHBWB:2021,barioni2025interpretable}, mathematically described by the time-delayed nonlinear Lang–Kobayashi (LK) equations~\cite{LK:1980}. 

Previous studies of evanescently coupled semiconductor laser arrays~\cite{winful1988stability,winful:1990} demonstrated that phase-locked states are readily destabilized above threshold, with the onset of instability depending critically on the ratio of carrier to photon lifetimes. For an array of fully coupled external cavity semiconductor lasers with spatially decaying coupling, two recent theoretical studies~\cite{NHBWB:2021,barioni2025interpretable} proposed achieving and enhancing synchronization by exploiting disorder. Specifically, Ref.~\cite{NHBWB:2021} demonstrated that synchronization can be promoted when time-delay-induced disorder compensates for intrinsic frequency disorder, while Ref.~\cite{barioni2025interpretable} showed that frequency synchronization can be stabilized by introducing intermediate random mismatches among otherwise identical lasers, including frequency, coupling, and amplitude-phase coupling disorders. In both studies, the disorders were treated as tunable parameters. While it is possible to experimentally tune the frequency disorder of semiconductor lasers by changing either electrical current or optical pump power for individual lasers, this requires precise control that may be experimentally challenging.

\begin{figure} [ht!]
\centering
\includegraphics[width=\linewidth]{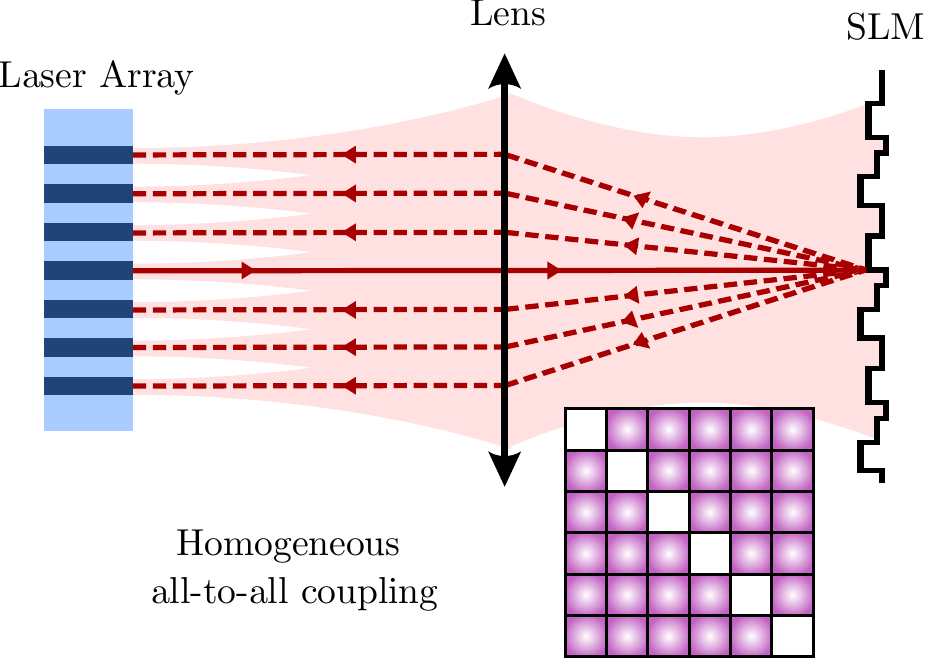}
	\caption{ Illustration of a network of coupled semiconductor diode lasers. Experimentally, coupling is implemented using a spatial light modulator (SLM) placed at the back focal plane of a collimation lens, with the laser array located at the front focal plane. Lasers at different transverse positions are coupled through gratings of corresponding periods on the SLM, introducing a delay determined by the round-trip time $\tau$ between the array and the SLM. The intrinsic lasing frequencies of the individual lasers are randomly detuned due to unavoidable fabrication-induced heterogeneity.}
\label{fig:schematic}
\end{figure}

Frequency disorder in semiconductor laser networks~\cite{NHBWB:2021,barioni2025interpretable,mikaberidze2025emergent,ye2025optimal} is unavoidable, primarily arising from manufacturing imperfections~\cite{mu2022extreme,zhang2024single}. Such disorder disrupts synchronization and limits the achievable coherence and output power. Although chaotic dynamics in semiconductor lasers have been successfully exploited for applications such as secure communication~\cite{ASLALCFGOMPS:2005,URGOR:2005}, random number generation~\cite{UAIHNSOKSYYD:2008}, and LiDAR/radar systems~\cite{LL:2004a,LL:2004b,CCCPTL:2018}, they can be detrimental in experiments and systems that require coherent, high-power output. In such contexts, stable steady-state synchronization is the preferred operating regime.
Recent advances in optical control have enabled the engineering of coupling topologies~\cite{brunner:2015}, for example, all-to-all coupling implemented via a spatial light modulator (SLM), as shown in Fig.~\ref{fig:schematic}, which can also be realized without an SLM~\cite{BMHPR:1988,KVM:2000,ABS:2016}. In practice, however, the total coupling strength, which reflects the experimental coupling cost, is often constrained. This raises a fundamental question: for a given configuration of frequency disorder and within the weak-coupling regime, can one identify an optimal coupling strength that maximizes steady-state synchronization in disorder-mediated synchronization resonance across the entire network?

This paper provides an affirmative answer. We demonstrate that, for fixed frequency disorder, the coupling strength in a homogeneous, all-to-all coupled laser network can be tuned to achieve high-quality steady-state synchronization. In the weak-coupling regime, the global coupling strength is much smaller than the statistical spread (standard deviation) of the frequency disorder. Qualitatively, steady-state synchronization arises from the intricate interplay among frequency disorder, coupling effects, and dynamical stability in the optical fields. In the absence of coupling, the free-running lasers remain in steady states with different intrinsic frequencies, and this frequency disorder prevents synchronization. Increasing the global coupling strength within the weak-coupling regime can gradually pull all different frequencies toward the same final dynamical frequency but eventually drives the system toward chaotic dynamics, which are undesirable, as shown in Fig.~\ref{fig:all2all}. Consequently, as the coupling strength increases, steady-state synchronization reaches its maximum just before the onset of chaos, representing a trade-off among disorder, coupling, and stability. A remarkable finding is that, the optimal coupling strength for maximizing steady-state synchronization scales inversely with the network size, making it possible to synchronize a large number of semiconductor lasers in the weak coupling regime. This leads to a linear scaling of the total coupling cost with the number $M$ of lasers, which is much more desirable than the $M^2$ scaling if the critical coupling coefficient were independent of $M$.  

To explain our findings, we develop a theoretical framework that recasts the delayed phase dynamics as a gradient flow on an effective thermodynamic potential, where steady-state solutions correspond to local minima of the potential. This framework elucidates the observed scaling of the optimal coupling with system size. Although it has been qualitatively recognized in the laser community that coupling must be neither too weak nor too strong to achieve synchronization, a systematic study and quantitative understanding in semiconductor laser networks have been lacking until now.

\section{Synchronization in complex networks and stochastic resonance} \label{sec:syn_SR}

Parallel to the emergence of modern network science about 25 years ago, synchronization in complex networks has been extensively investigated~\cite{ADKMZ:2008}, owing to the ubiquity of networked structures in natural and engineered systems and the fundamental importance of synchronous dynamics for their function. Early studies revealed that small-world networks, characterized by short path lengths, are more synchronizable than regular lattices of comparable size~\cite{BP:2002}, whereas structural heterogeneity typically hinders synchronization~\cite{NMLH:2003}. Later work demonstrated that heterogeneous networks with properly weighted links can outperform small-world and random networks in terms of synchronizability~\cite{WLL:2007}. The onset of chaotic phase synchronization in networks of coupled heterogeneous oscillators was also explored~\cite{RTHL:2012}, and the interplay between network symmetry and synchronization was elucidated~\cite{PSHMR:2014,SPHMR:2016,HZRM:2019}.

In general, the transition from desynchronization to synchronization in complex networks is continuous, which is characteristic of a second‑order phase transition, where the synchronization error or order parameter evolves smoothly through the critical point. However, studies of coupled phase oscillators have revealed that, when network couplings are weighted according to the oscillators’ natural frequencies, the transition can become abrupt and discontinuous, indicative of a first‑order, or explosive, synchronization phenomenon~\cite{GGAM:2011,ZPSLK:2014,BAGLLSWZ:2016,LZTL:2022}. This behavior demonstrates the strong influence of network structure on collective dynamics. It has further been shown mathematically that temporally varying, random coupling can promote synchronization~\cite{ZGLLL:2019}. Together, these results underscore how coupling architecture, structural heterogeneity, and temporal modulation shape collective behavior and connect naturally to other lines of research where noise or disorder enhances coherent response through resonance mechanisms.

Stochastic resonance, coherence resonance, and disorder‑induced resonance are three distinct yet related phenomena in nonlinear and stochastic dynamical systems. Stochastic resonance~\cite{benzi1981mechanism,gammaitoni1998stochastic} occurs when a system subject to noise is driven by a weak periodic signal; at an optimal noise amplitude, the response to the periodic driving, quantified by the signal-to-noise ratio, is maximized. Coherence resonance~\cite{pikovsky1997coherence,pisarchik2023coherence} arises even without external driving, when noise alone induces a timescale that matches the intrinsic timescale of the system, leading to enhanced temporal regularity. In both cases, the key ingredient is time-dependent stochastic forcing.

In spatially extended dynamical systems, time-independent disorder can play an analogous role. When such a system is driven by a weak periodic signal, an optimal level of spatial disorder can maximize its response, a phenomenon known as disorder-induced resonance~\cite{tessone2006diversity,liu2024quenched}. For instance, in a network of bistable oscillators, random parameter disorder across oscillators can facilitate collective dynamics: a subset of oscillators first responds to the weak periodic drive as coupling increases, acting as ``seeds'' that subsequently entrain the rest of the network. Although static spatial disorder may be viewed as a limiting case of time-dependent noise, the underlying mechanism of disorder-induced resonance differs fundamentally from that of stochastic or coherence resonance.

In all three forms of resonance, tuning the noise or disorder amplitude to an optimal level maximizes a specific measure of system performance. The setting of our study is fundamentally different. The intrinsic frequency disorder in semiconductor lasers is inherent to the devices and fixed. Moreover, our focus is on steady-state synchronization, whereas studies of stochastic, coherence, or disorder-induced resonances typically involve oscillatory or complex temporal dynamics. The motivation for our work is strictly experimental and application-oriented: in practice, spatial disorder is unavoidable and fixed, and the ultimate goal is to realize highly coherent, high-power, steady-state laser emission.

\section{Results} \label{sec:results}

A network of $M$ coupled diode lasers, each operating in a single longitudinal 
and transverse mode at fixed polarization with frequency disorders, is described 
by the LK equations~\cite{LK:1980,NHBWB:2021}:
\begin{align} \nonumber
	\frac{d E_{i}(t)}{dt}&=\frac{1+i\alpha}{2}\left(g\frac{N_{i}(t)-N_{0}}{1+s|E_{i}(t)|^{2}}-\gamma\right)E_{i}(t) \\ \notag &  + i\Delta_i E_{i}(t) + e^{-i\omega_{0}\tau}\kappa\sum_{j\neq i}E_{j}(t-\tau), \\ \label{eq: LK}
    \frac{dN_{i}(t)}{dt}&=J_{0}-\gamma_{n}N_{i}(t)-g\frac{N_{i}(t)-N_{0}}{1+s|E_{i}(t)|^2}|E_{i}(t)|^2,
\end{align}
where $E_{i}(t)$ is the complex electric field of the $i$th laser, $N_{i}(t)$ is 
the carrier number governed by the pump rate $J_0 = 4 J_{th}$ with 
$J_{th} = \gamma_{n}(N_{0}+\gamma/g)$ being the single-laser pump rate at 
threshold, $\gamma_{n}$ is the carrier loss rate, $N_0$ is the carrier number at 
transparency, $\gamma$ is the cavity loss rate, $g$ is the differential gain 
coefficient, and $\tau$ is the time delay due to the external cavity. The typical 
experimental parameter values are $\gamma_{n}=0.5\,\textnormal{ns}^{-1}$, 
$N_{0}=1.5\times 10^{8}$, $\gamma=500\,\textnormal{ns}^{-1}$, and 
$g=1.5\times\,10^{-5}\,\textnormal{ns}^{-1}$. Other experimental parameter 
values are: the amplitude–phase coupling (linewidth enhancement) factor $
\alpha=5$, gain saturation coefficient $s=2\times 10^{-7}$, and external time 
delay $\tau= 3\,\textnormal{ns}$. Random frequency disorders are modeled as 
$\Delta_i = \sigma_{\Delta}\,\mathcal{N}(0,1)$, where $\mathcal{N}(0,1)$ is a 
Gaussian random variable with zero mean and unit variance. The natural angular 
frequency reference $\omega_0$ is chosen such that $\omega_0\tau=2N\pi$ with 
$N$ selected to be closest to the mean of all natural angular frequencies. The 
frequency disorder values are ordered as 
$\Delta_1\leq\Delta_2\leq\ldots\leq\Delta_M$, with 
$\sigma_{\Delta}=14\,\textnormal{rad/ns}$ and $M=24$. 

The laser field is expressed as $E_i(t)=r_i(t) e^{i\Omega_i(t)}$. Synchronization, encompassing the same amplitude, frequency, and phase, can be quantified by $\langle S\rangle=\langle |\sum_{i=1}^M E_i(t)|^2/[M\sum_{i=1}^M|E_i(t)|^2]\rangle\in[0,1]$, where a larger value indicates stronger synchronization. If only frequency and phase synchronization are of interest, then $\langle S\rangle$ reduces to $\langle R^2\rangle=\langle |\sum^{M}_{i=1}e^{i\Omega_i(t)}|^2/M^2\rangle$ since all lasers share the same amplitude $r_0(t)$. The Kuramoto model is a reduced phase-dynamics model derived from the LK equation, so $\langle R^{2}\rangle$ is its standard choice. Because the LK equations include nonzero amplitude–phase coupling, it is preferable to use $\langle S\rangle$ so that amplitude synchronization is also taken into account. The all-to-all coupling configuration is described by $K_{ij} = \kappa(1 - \delta_{ij})$, with $\kappa$ and $\delta_{ij}$ being the global coupling strength and Kronecker delta, respectively. The self-feedback coupling (the diagonal elements of the coupling matrix) is set to zero to avoid dynamical instabilities. Without frequency disorders, complete synchronization ($\langle S\rangle=1$) occurs in both the weakly ($\kappa\ll\sigma_{\Delta}$) and strongly ($\kappa\gg\sigma_{\Delta}$) coupling regime (Appendix~\ref{appendix_a}). 

\begin{figure} [ht!]
\centering
\includegraphics[width=\linewidth]{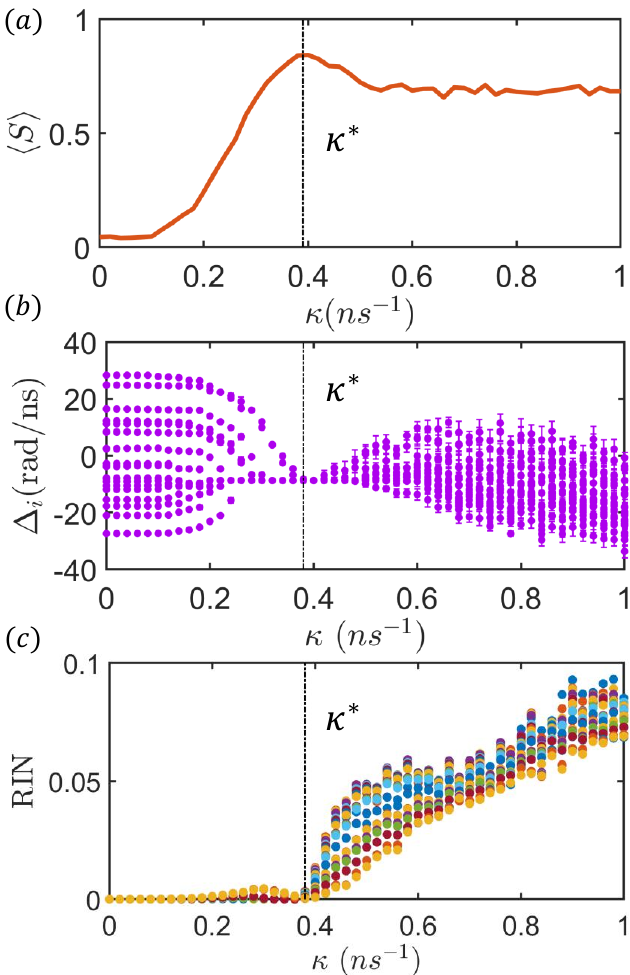}
\caption{Optimized steady-state synchronization with fixed frequency disorder in a network of $M=24$
semiconductor diode lasers. (a) Maximization of synchronization measures
$\langle S\rangle$ by an optimal coupling strength 
$\kappa^*$. (b) Average short-term final frequency detuning of the individual 
lasers versus $\kappa$, where the error bars represent the standard deviation 
calculated over a moving time window of size $3\tau$ with a step size of 
$0.3\tau$ for $t\in [50,100]\rm ns$. 
(c) Mean-square value of the normalized intensity fluctuations, $(I_i(t)-\langle I_{i}(t)\rangle)/\langle I_{i}(t)\rangle$, as a function of the coupling parameter $\kappa$, serving as a time-domain measure of the relative intensity noise (RIN) of individual lasers over a $50\,\rm ns$ time window. Different colors correspond to different lasers.}
\label{fig:all2all}
\end{figure}

\begin{figure*} [ht!]
\centering
\includegraphics[width=\linewidth]{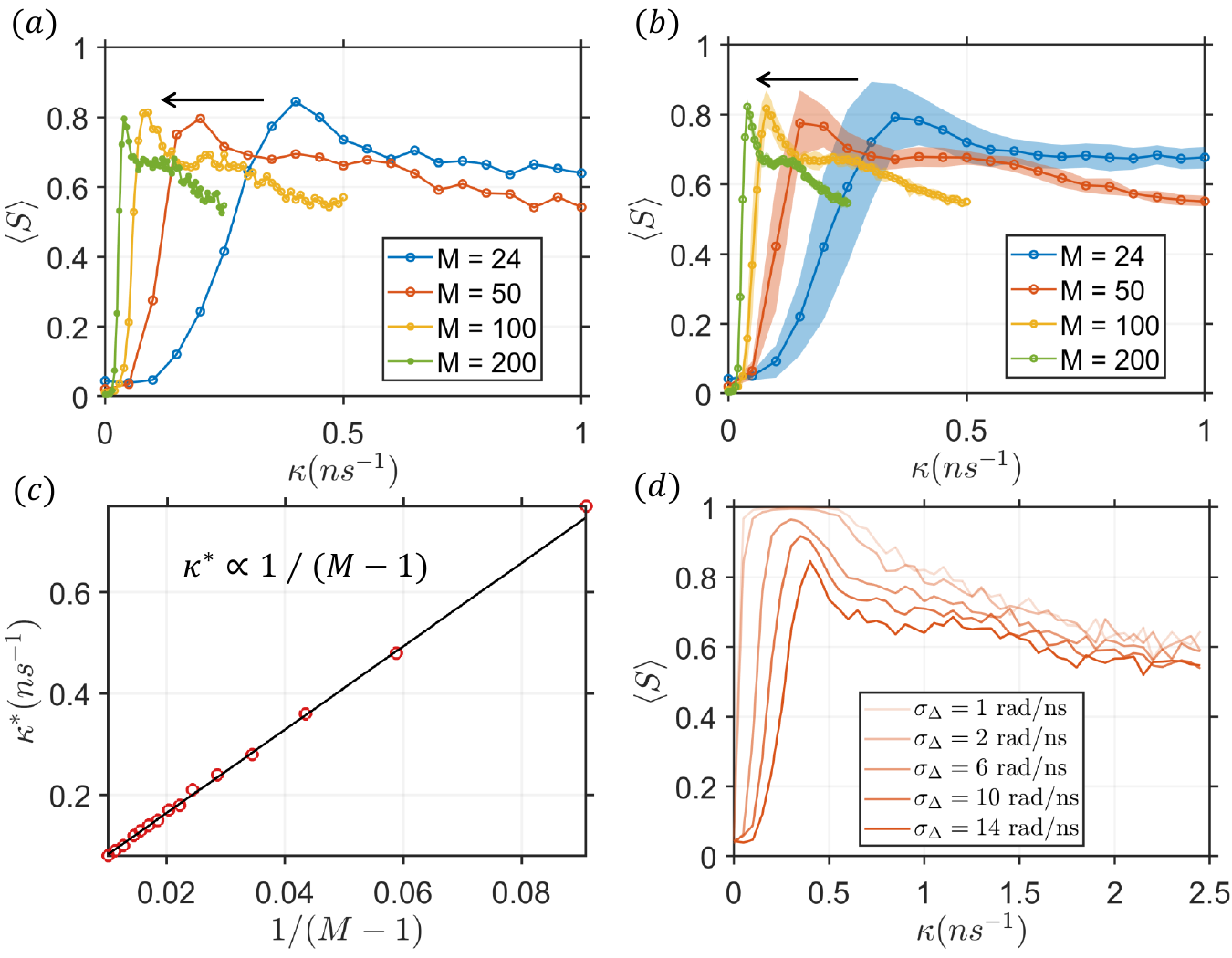}
\caption{Robustness and size scaling of optimized steady-state synchronization. The frequency 
disorders are independently sampled from a Gaussian distribution 
$\sigma_{\Delta} \mathcal{N}(0,1)$ with the standard deviation $\sigma_{\Delta}$. 
(a) Optimizing steady-state synchronization for different number $M$ of lasers for 
$\sigma_{\Delta} = 14\,\textnormal{rad/ns}$. (b) Statistical fluctuations
of the optimized steady-state synchronization, where for each value of $M$, ten independent realizations of the frequency disorder for $\sigma_{\Delta} = 14\,\textnormal{rad/ns}$ are
tested. The four curves in their respective statistical clouds are the averages 
and the shaded areas indicate the corresponding standard deviations. (c) Location 
of the optimized peak, $\kappa^*$, indicated by the averaged peak position over 
ten frequency disorder realizations. The resulting $\kappa^*$ values exhibit an 
inverse relationship with the system size: $\kappa^* \propto 1/(M-1)$, 
demonstrated for $M = [12,18,24,30,36,42,46,50,55,60,65,70,80,90,100]$. 
(d) Effect of frequency disorder strength as characterized by $\sigma_{\Delta}$ 
on the steady-state synchronization for $M=24$. For small values of $\sigma_{\Delta}$, 
the peak value of $\langle S\rangle$ can reach unity as in the corresponding 
disorder-free system.}
\label{fig:resonance_scaling}
\end{figure*}

When frequency disorder is present, synchronization resonance and optimized steady-state synchronization emerge in the weak-coupling regime ($\kappa\ll\sigma_{\Delta}$), with the optimal coupling $\kappa^{*} \in [0,0.5]\,\textnormal{ns}^{-1}$ for $M\geq 24$. The synchronization measure attains $\langle S\rangle >0.8$ (approaching unity) over the range $\sigma_{\Delta}\in[1,14]\,\textnormal{rad/ns}$,  independent of the number of lasers $M$, as shown in Figs.~\ref{fig:all2all} and \ref{fig:resonance_scaling}.
For example, for $M=24$ lasers, we have $\langle S \rangle_{\rm max}  \approx 0.84$ 
for $\kappa^* \approx 0.4~\textnormal{ns}^{-1}$, as shown in 
Fig.~\ref{fig:all2all}(a). Figures~\ref{fig:all2all}(b) and \ref{fig:all2all}(c) 
display the fitted short-term frequencies and the fluctuations of the normalized 
intensities for $M = 24$, respectively. Our computations revealed that, in the 
presence of frequency disorders, the strong synchronization $\langle S\rangle >0.8$ achieved in the weak coupling regime, is in fact steady-state synchronization, where the individual lasers 
maintain sinusoidal, nearly synchronous oscillations. This type of synchronization 
is desired in applications. 
In the strong‑coupling regime ($\kappa\gg\sigma_{\Delta}$), chaotic dynamics emerge~\cite{nair:2019almost,nair:2018phase,NHBWB:2021}, but these are not relevant to steady‑state synchronization.

Heuristically, the underlying mechanism for synchronization resonance and optimized steady-state synchronization, which occurs near the boundary between the steady‑state and chaotic dynamical regimes in Fig.~\ref{fig:all2all}, can be described as follows. For $\kappa = 0$, the presence of frequency disorder prevents synchronization under steady‑state dynamics. At small $\kappa$, the coupling is too weak to compensate for the frequency detuning, leading to only marginal improvement in synchronization. As $\kappa$ increases, synchronization is progressively enhanced as long as the laser dynamics remain regular (non‑chaotic). This trend ceases once irregular dynamics emerge, as shown in Figs.~\ref{fig:all2all}(b) and \ref{fig:all2all}(c), resulting in a subsequent decrease in the synchronization measure $\langle S\rangle$ with further increases in $\kappa$. For sufficiently strong coupling $\kappa \gg \sigma_{\Delta}$, the disruptive effect of disorder on synchronization becomes negligible; however, this regime is beyond the scope of the present study. The onset of chaos stems from the combined influence of stronger coupling and nonzero amplitude-phase coupling $\alpha$. In the absence of amplitude-phase coupling, the dynamics revert to a steady state at the stronger coupling compared with the weak-coupling regime ($\kappa\ll\sigma_{\Delta}$) (Appendix~\ref{appendix_b}). Qualitatively, the optimized steady-state synchronization emerges from the interplay among frequency disorder, coupling, and dynamical stability.

Is the optimized steady-state synchronization robust as the laser network becomes 
larger? Figure~\ref{fig:resonance_scaling} provides an affirmative answer. As 
the network size $M$ increases, the optimal coupling value $\kappa^*$ decreases, 
but the optimized steady-state synchronization persists, as shown in Fig.~\ref{fig:resonance_scaling}(a). This can be understood by noting that the total coupling per laser 
$\sum_{j} K_{ij} = \kappa (M-1)$ is assumed to be fixed as in a realistic 
experimental setting. If approximately the same amount of total coupling per laser 
is required for the optimized steady-state synchronization to arise, increasing $M$ will naturally result in a reduced $\kappa^*$ value. As $M$ increases, the statistical variation of the 
frequency disorder becomes well behaved, leading to narrower fluctuations in 
the synchronization optimization, as shown in Fig.~\ref{fig:resonance_scaling}(b). 
The height of the optimized peak in $\langle S \rangle$ with increasing $M$ 
remains approximately constant, due to the fact that,
at the optimized peak, the total coupling strength per laser remains constant, 
thereby preserving the steady-state solution and sustaining the same level of 
synchronization, as explained by our theory below. Quantitatively, the dependence 
of $\kappa^*$ on $M$ follows the scaling relation: $\kappa^{*} \propto 1/(M-1)$, 
as shown in Fig.~\ref{fig:resonance_scaling}(c). For $M=2$ ($M=24$) coupled lasers, 
the optimal coupling is $\kappa^{*}\approx 8~\mathrm{ns^{-1}}$ 
($\kappa^{*}\approx 0.4\,{\rm ns^{-1}}$) [Fig.~\ref{fig:resonance_scaling}(c)]. 
The scaling law indicates that increasing the number of lasers significantly reduces the required coupling strength per link ($\kappa^{*}$), whereas the optimal total coupling cost, $\sum_{ij}K^{*}_{ij}\propto M$, grows linearly with $M$, consistent with our physical theory presented below.
In addition, as the extent of frequency disorder measured by the 
standard deviation $\sigma_{\Delta}$ of the frequency distribution is reduced, 
synchronization is enhanced, as shown in Fig.~\ref{fig:resonance_scaling}(d).

To explain the optimized steady-state synchronization in Figs.~\ref{fig:resonance_scaling}(a–d), 
we develop a physical theory by expanding the LK equations about the steady state, 
yielding a generalized time-delayed Kuramoto system~\cite{yeung:1999} with an 
additional phase shift $\tan^{-1}\alpha+\omega_0\tau$ and an effective coupling 
enhanced by the amplitude–phase coupling $\sqrt{1+\alpha^2}$ (see Appendix~\ref{appendix_c} for more details). Introducing the delayed phase differences 
$\eta_i = \Omega_i(t)-\Omega_i(t-\tau)$, the dynamics can be recast as a gradient 
flow governed by an effective thermodynamic potential $U(\eta_i)$, which 
encapsulates the collective coupling influence of the other lasers on laser $i$. 
In the absence of frequency disorder ($\Delta_i=\Delta_j=0$) with all lasers 
oscillating at the reference frequency $\omega_0$, the coupling configuration 
$K_{ij}=\kappa(1-\delta_{ij})$ renders the potential identical across the lasers:
\begin{align} \label{eq:potential}
    U(\eta(t))=\eta^{2}(t)-2\tau\sqrt{1+\alpha^2}&k^{\textnormal{in}}\cos\big[\eta(t)\notag\\
    &+\tan^{-1}\alpha+\omega_0\tau\big],
\end{align} 
where $k^{\rm in}=\sum_j K_{ij}=\kappa(M-1)$ is the intrinsic coupling strength 
received by each laser from the others. The local minima of $U(\eta)$ coincide 
for all lasers at $\eta^{*}=2\pi f_{\rm final}\tau$ that defines the synchronized 
frequency. The first quadratic term enforces parabolic confinement, 
while the cosine modulation arises from delay $\tau$, amplitude-phase coupling 
parameter $\alpha$, and intrinsic coupling $k^{\rm in}$. For weak effective 
coupling $\mathbb{K}\equiv\tau\sqrt{1+\alpha^2}\kappa(M-1)\ll \eta^{2}$, the 
modulation is negligible and the system is in a unique synchronized steady 
state. In contrast, stronger coupling generates multiple 
local minima leading to multistability that drives the system into chaotic 
regimes for $\alpha\ne 0$ (Appendix~\ref{appendix_c}).

\begin{figure*} [ht!]
\centering
\includegraphics[width=0.86\linewidth]{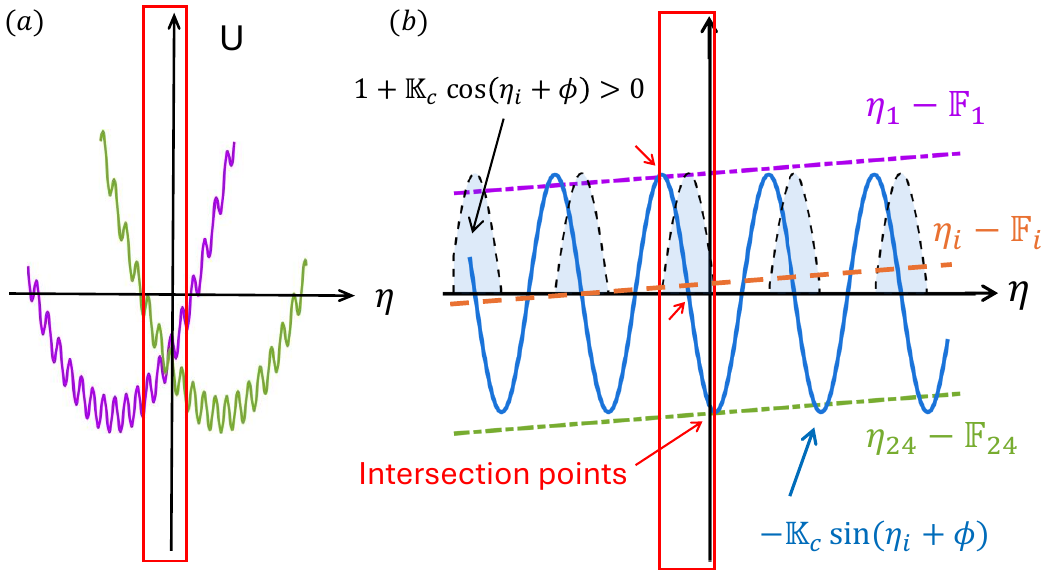}
\caption{Schematic of local-minima selection, corresponding to the steady state 
of each laser. (a) Effective thermodynamic potential comprising a parabolic term, 
a shifted global minimum $-2\tau\Delta_i\eta_i(t)$, and a coupling-induced cosine 
component. The red rectangle marks the near-overlapping regime that yields nearly 
synchronized final frequencies. (b) Illustration of steady-state selection from 
the local minima of the effective potential landscape for different lasers at 
the critical effective coupling $\mathbb{K}_c$.}
\label{fig:theory}
\end{figure*}

In the system governed by a smooth quadratic (parabolic) potential without 
coupling, as indicated by Eq.~(\ref{eq:potential}), once the coupling-induced 
cosine oscillation is excluded, frequency disorder ($\Delta_i \ne \Delta_j$) 
contributes an additional term $-2\tau \Delta_i \eta_i(t)$ to the thermodynamic 
potential $U(\eta_i(t))$. As a result, the global minima of the individual 
lasers become substantially different, hindering synchronization. However, 
for sufficiently strong coupling, the potential landscape is reshaped by the 
cosine term in Eq.~(\ref{eq:potential}), producing multiple local minima 
separated by high barriers while leaving the distant global minima unchanged, 
as shown in Fig.~\ref{fig:theory}(a) for two distinct initial values of detuning 
$\Delta_i$. For a fixed effective coupling $\mathbb{K}$, the height of the 
barriers separating the local minima decreases monotonically with increasing 
$\eta$ from the global minimum. In the near-overlapping regime, e.g., the red 
rectangle in Fig.~\ref{fig:theory}(a), the proximity of these local minima 
leads to nearly synchronized final frequencies. This mechanism enables 
enhanced frequency and phase locking, even though the system ultimately 
settles into states that are not global minima.

More specifically, the local minima of the potential are determined by the 
conditions $dU/d\eta_i=0$ and $d^2U/d\eta^2_{i}>0$, leading to the constraints 
$\eta_i - \mathbb{F}_i = -\mathbb{K}\sin(\eta_i+\phi)$ and 
$1+\mathbb{K}\cos(\eta_i+\phi)>0$, respectively, where 
$\mathbb{F}_i\equiv \tau\Delta_i$ denotes the effective detuning and 
$\phi\equiv\textnormal{mod}(\tan^{-1}\alpha + \omega_0\tau,2\pi)\approx 0.4\pi$. 
Given the critical effective coupling $\mathbb{K}_c$, the intersections between 
the straight lines $\eta_i - \mathbb{F}_i$ and the sinusoidal curves 
$-\mathbb{K}_c\sin(\eta_i+\phi)$ that satisfy $d^{2}U/d\eta^{2}_{i}>0$ define 
the local minima of the potential, i.e., the steady-state solutions $\eta_i$ 
for each laser $i$, as illustrated in Fig.~\ref{fig:theory}(b). The corresponding 
local minima $\eta_i$ determine the final frequencies of the lasers. For small 
coupling $\mathbb{K}\ll\mathbb{K}_c$, laser frequencies remain widely separated, 
whereas large coupling $\mathbb{K}\gg\mathbb{K}_c$ induces multistability and 
chaos via unstable dynamical invariant sets. There exists a critical effective 
coupling determined by $\mathbb{K}_{\rm c} = \max(|\mathbb{F}_i|)$, at which 
all lasers attain their final frequencies within the same regime satisfying 
$1+\mathbb{K}\cos(\eta_i+\phi)>0$, ensuring the local minimum condition 
$d^2U/d\eta^2_i>0$, as highlighted by the red square in Fig.~\ref{fig:theory}(b). 
In this case, the maximum frequency difference is determined by the width of the 
regime where $d^{2}U/d\eta_i^{2}>0$. The onset of such solutions defines the 
critical effective coupling with the resulting states referred to as 
near-synchronized steady states. Below this threshold, only a subset of lasers 
achieve closely aligned final frequencies within the same 
$1+\mathbb{K}\cos(\eta+\phi)>0$ regime.

Beyond this critical effective coupling, the final frequency differences 
diminish and multiple stable yet distinct near-synchronized steady states emerge. 
As their stability weakens, synchronization begins to saturate, giving rise to 
periodic or quasiperiodic phase dynamics. As the coupling strength increases 
further, the nontrivial amplitude-phase coupling in the LK equation (Appendix~\ref{appendix_c}) 
can trigger chaos~\cite{winful1988stability,winful:1990}, thereby weakening
synchronization. The onset of chaos defines the optimized peak at $\kappa^{*}$, 
as shown in Fig.~\ref{fig:all2all}, which typically occurs slightly above the 
critical coupling strength, as $\kappa^{*}>\kappa_c$, where $\kappa_c$ is 
determined from the critical effective coupling $\mathbb{K}_c$. In contrast, 
for $\alpha=0$, the synchronization resonance and optimized steady-state synchronization in the weak coupling vanish: synchronized steady states appear only at the stronger coupling, while for weak coupling 
($\kappa\ll\sigma_{\Delta}$) the system is far away from synchronization (Appendix~\ref{appendix_c}).

More specifically, for the frequency disorder $\Delta_i=14\times\mathcal{N}(0,1)$ 
rad/ns, the maximum effective detuning is 
${\rm \max}(|\mathbb{F}_i|)=\tau|\Delta_0|\approx 90$. To compensate for it, the 
effective coupling $\mathbb{K}^{*}$ must exceed $\mathbb{K}_c=\tau|\Delta_0|$, 
giving $\kappa^{*}>\kappa_c=|\Delta_0|/[(M-1)\sqrt{1+\alpha^2}]\approx 0.26\,{\rm ns}^{-1}$ 
for $M=24$, which is consistent with Fig.~\ref{fig:resonance_scaling}. This 
relationship reveals the scaling $\kappa^{*}\propto 1/(M-1)$, thereby explaining 
the numerical results in Fig.\ref{fig:resonance_scaling}(c). Moreover, the 
critical coupling coincides with the necessary condition of the classical 
Kuramoto model\cite{chopra:2009,jadbabaie:2004,dorfler:2011} in the limit of 
vanishing amplitude-phase coupling and time delay. 

\section{Discussion} \label{sec:discussion}

To summarize, we demonstrate that disorder-mediated synchronization resonance and optimal steady-state synchronization can be achieved in the weak-coupling regime of semiconductor diode-laser networks with fixed frequency disorder, and we develop a physical theory that explains both its origin and the associated size-scaling behavior. Remarkably, the peak synchronization level is independent of the number of lasers, while the required coupling cost scales linearly with system size, ensuring robust synchronization even in large arrays. The emergence of this optimal state near the boundary between steady-state and chaotic regimes results from the competition among frequency disorder, coupling effects, and dynamical stability. More broadly, the complex interplay of amplitude and phase dynamics together with time-delay effects makes optimal steady-state synchronization a robust and experimentally observable phenomenon across a wide range of network sizes.

Beyond edge-emitting diode lasers (described by the LK equations), our results can be extended to vertical-cavity surface-emitting lasers and other classes of nonlinear oscillators. In general, variations in intrinsic time constants can influence the onset of chaotic behavior and thereby modify the optimal coupling strength. Overall, the present findings extend beyond the traditional Kuramoto paradigm and establish semiconductor laser networks as a promising platform for investigating disorder-enhanced collective dynamics.

\section*{Acknowledgement}

This work was supported by the Office of Naval Research under Grant No.~N00014-24-1-2548.

\section*{Data Availability}

All data are simulation data and are available upon request.

\appendix

\section{Complete synchronization without frequency disorder and $\alpha$-induced chaotic regimes} \label{appendix_a}

\begin{figure*} [ht!]
\centering
\includegraphics[width=0.9\linewidth]{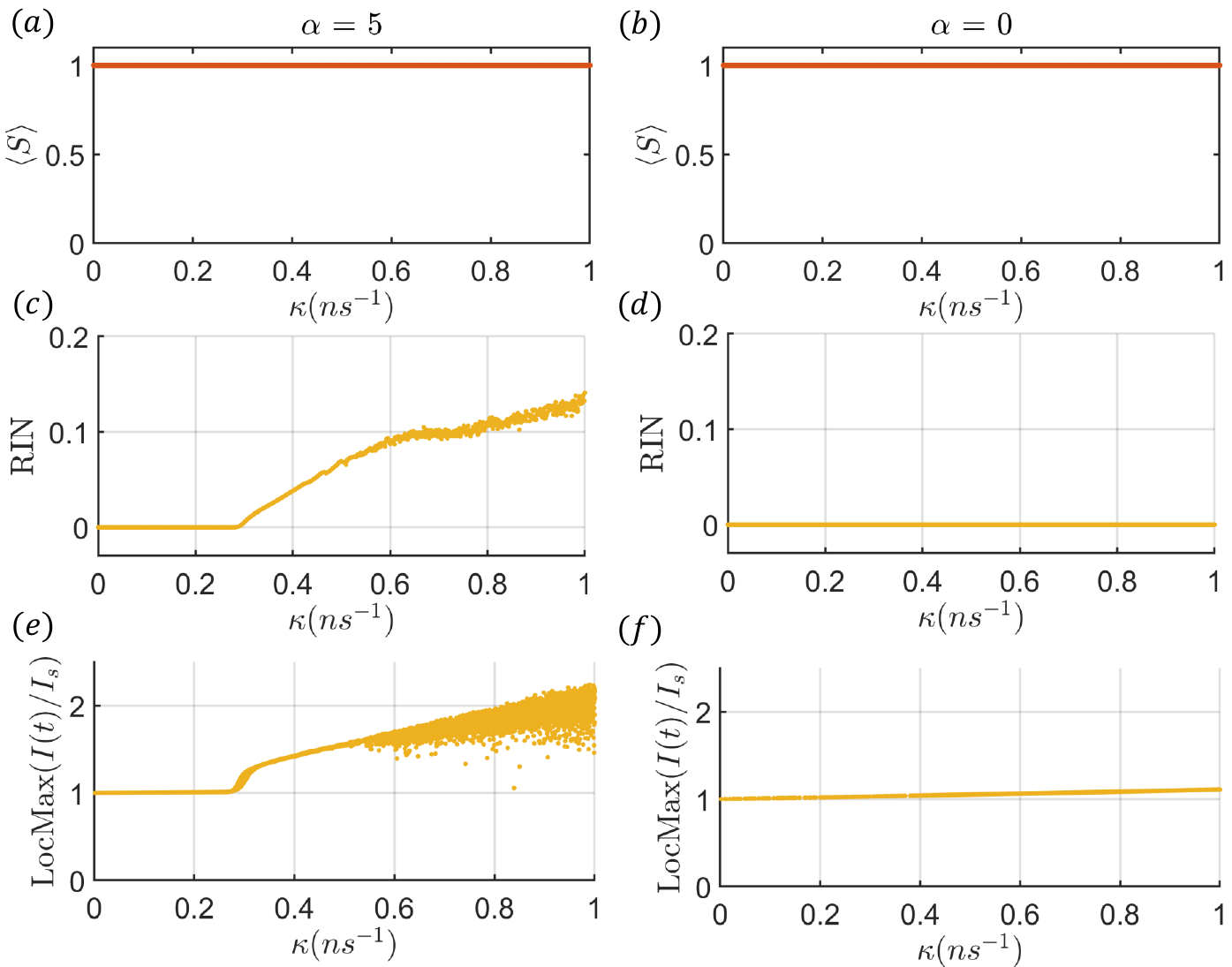}
\caption{Without frequency disorder, homogeneous all-to-all coupling yields complete synchronization, while nonzero amplitude–phase coupling $\alpha$ drives the system into chaotic regimes as the coupling increases. Panels (a,c,e) and (b,d,f) show, for $M=24$, the synchronization measure $\langle S \rangle$, a time-domain measure of RIN, and the local maxima of the normalized intensity for each laser $i$, as functions of $\kappa$ for $\alpha=5$ and $\alpha=0$. The time window is $[50,100]\,{\rm ns}$. Complete synchronization means that all $M=24$ lasers share the same RIN and the same local maxima of the normalized intensity.}
\label{fig:supp_alpha_disorder_free}
\end{figure*}

In the absence of frequency disorder ($\Delta_i=\Delta_j=0$) with all lasers oscillating at the reference frequency $\omega_0$, homogeneous all-to-all coupling $K_{ij}=\kappa(1-\delta_{ij})$ yields complete synchronization with $\langle S\rangle=1$ for arbitrary $\kappa\in[0,1]\,\rm ns^{-1}$. The underlying dynamics, however, depend on the amplitude–phase coupling $\alpha$. For $\alpha=0$, the system can be reduced approximately to pure phase dynamics governed by the Kuramoto model, leading to steady-state synchronization independent of $\kappa$. By contrast, a nonzero $\alpha$ in the Lang–Kobayashi (LK) equations drives the system into chaos once the coupling enters the stronger coupling regime with $\kappa \gtrsim 1\,{\rm ns^{-1}}$~\cite{winful1988stability,winful:1990}.

\section{The $\alpha$–driven optimization of steady-state synchronization in the weak-coupling regime} \label{appendix_b}

\begin{figure*} [ht!]
\centering
\includegraphics[width=0.9\linewidth]{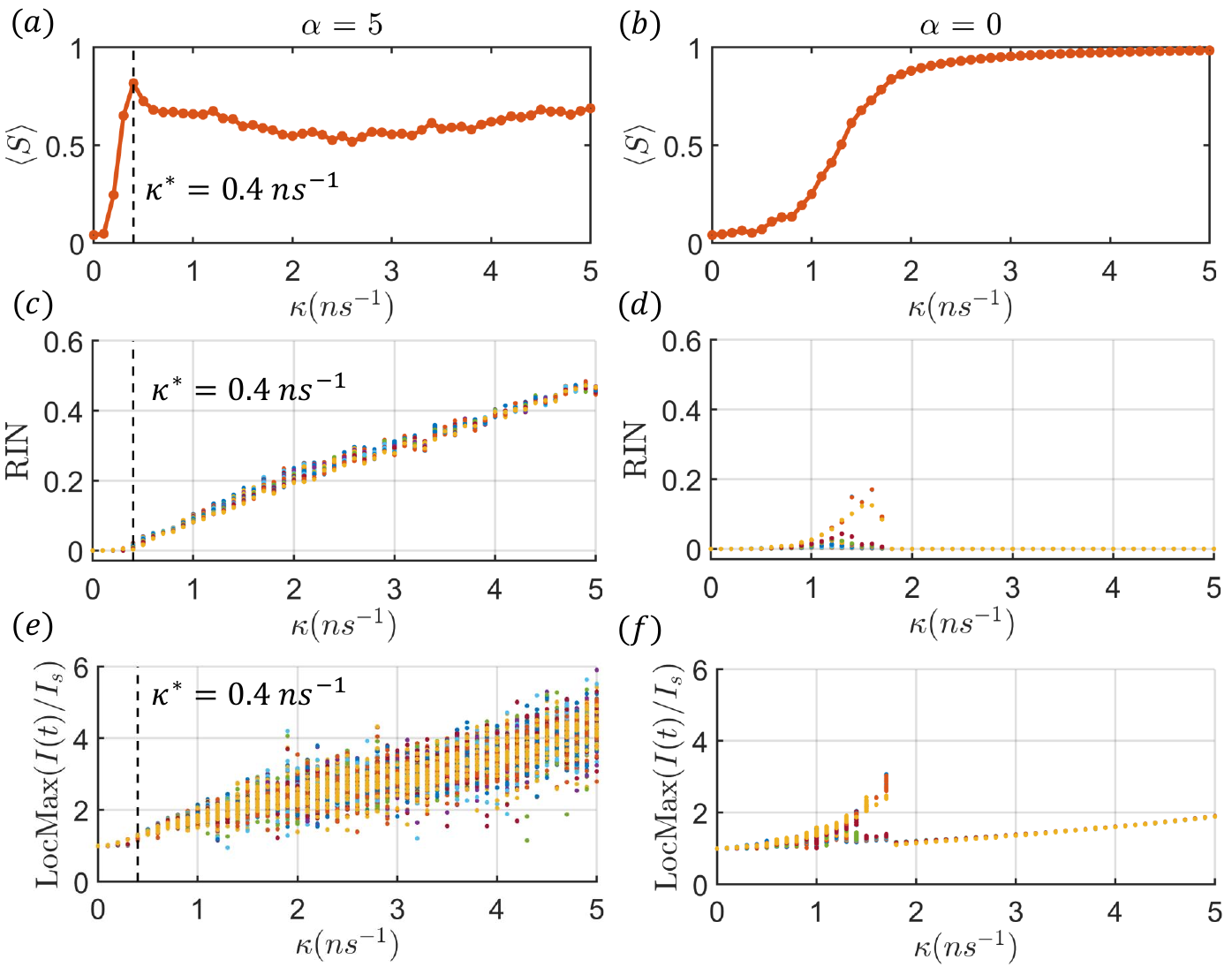}
\caption{Under fixed frequency disorder, the amplitude–phase coupling induces chaotic dynamics at stronger coupling than in the weak-coupling regime $\kappa\ll\sigma_{\Delta}$, leading to the emergence of optimized steady-state synchronization—an effect absent in the purely phase-based dynamics of the classical Kuramoto model. Panels (a,c,e) and (b,d,f) show, for $M=24$, the synchronization measure $\langle S \rangle$, a time-domain measure of RIN, and the local maximum of the normalized intensity for each laser $i$, as functions of $\kappa$ for $\alpha=5$ and $\alpha=0$. The time horizon is $[150,200]\,{\rm ns}$. Different colors represent different lasers.}
\label{fig:supp_alpha}
\end{figure*}

In the main text, we develop the thermodynamic potential theory to analyze the structure of near-synchronized steady states. Based on the critical coupling strength, this framework explains both the scaling behavior of optimized steady-state synchronization at the optimal coupling value $\kappa^{*}$ and the independence of the synchronized peak $\langle S\rangle_{\rm max}$ from the number of lasers $M$. However, the specific optimal coupling $\kappa^{*}$—typically larger than the critical value $\kappa_c$—is determined by the onset of chaotic dynamics beyond the steady state. This chaotic behavior, induced by the amplitude–phase coupling factor $\alpha$ (see Fig.~\ref{fig:supp_alpha}), is the direct reason for the emergence of the optimized steady-state synchronization in the weak-coupling regime ($\kappa\ll\sigma_{\Delta}$). Consequently, at the stronger coupling, although numerous near-synchronized steady states exist in both the LK equation and the Kuramoto model, these states are unstable chaotic attractors in the LK system due to nontrivial amplitude–phase coupling, but remain stable fixed points in the pure phase dynamics of the Kuramoto model.

\section{Effective thermodynamic potential for the laser network}\label{sec:thermodynamic_potential} \label{appendix_c}

Starting from the coupled LK equation in Eq. (1) of the main text with the generalized coupling matrix $K_{ij}$, 
we express the electric field in the polar coordinates as 
$E_i(t)=r_i(t)e^{i\Omega_i(t)}$, where $r_i(t)$ and $\Omega_i(t)$ are the 
amplitude and phase, respectively. To separate the radial and angular components, 
we multiply both sides of the complex electric field equation by 
$e^{-i\Omega_i(t)}/r_i(t)$. The real and imaginary parts of the resulting 
expression correspond to the amplitude and phase dynamics, respectively. 
The resulting coupled LK equations in the polar 
coordinates~\cite{NHBWB:2021} is 
\begin{widetext}
\begin{align}
    \frac{\dot{r}_i(t)}{r_i(t)} &= \frac{1}{2}[G(N_i(t),r_i(t))-\gamma]+\sum^{M}_{j=1}K_{ij}\frac{r_j(t-\tau)}{r_i(t)}\cos[\Omega_{j}(t-\tau)-\Omega_i(t)-\omega_0\tau],\notag\\
    \dot{\Omega}_{i}(t) &= \frac{\alpha}{2}[G(N_i(t),r_i(t))-\gamma]+\Delta_i+\sum^{M}_{j=1}K_{ij}\frac{r_j(t-\tau)}{r_i(t)}\sin[\Omega_j(t-\tau)-\Omega_i(t)-\omega_0\tau],\notag\\
    \frac{\dot{N}_{i}(t)}{N_i(t)} &= \frac{J_0}{N_i(t)}-\gamma_n - \frac{G(N_i(t),r_i(t))r^2_{i}(t)}{N_i(t)},
\end{align}
\end{widetext}
where the gain function is defined as
$$G(N_i(t),r_i(t))=g\frac{N_i(t)-N_0}{1+s\,r^{2}_i(t)}.$$
Since we are only concerned with the steady state, we now expand the solutions around it:
\begin{equation}
	r_i(t)\approx r^{s}_i, \ \ N_i(t)\approx N^{s}_i, \ \ {\rm and} \ \ \frac{r_j(t-\tau)}{r_i(t)}\approx 1.
\end{equation}
Under these approximations, the LK equations reduce to:
\begin{widetext}
\begin{align}
    0 &= \frac{1}{2}[G(N^{s}_i,r^{s}_i)-\gamma]+\sum^{M}_{j=1}K_{ij}\cos[\Omega_j(t-\tau)-\Omega_i(t)-\omega_0\tau],\label{eq: radius}\\
    \dot{\Omega}_i(t) &= \frac{\alpha}{2}[G(N^{s}_i,r^{s}_i)-\gamma]+\Delta_i+\sum^{M}_{j=1}K_{ij}\sin[\Omega_j(t-\tau)-\Omega_i(t)-\omega_0\tau],\label{eq: angle}\\
   0 &= \frac{J_0}{N^{s}_i}-\gamma_n - \frac{G(N^{s}_i,r^{s}_i)(r^{s}_{i})^2}{N^{s}_i}.
\end{align}
\end{widetext}
Substituting the gain function from Eq.~\eqref{eq: radius} 
into Eq.~\eqref{eq: angle} eliminates the gain term in Eq.~\eqref{eq: angle},
resulting in a phase equation that no longer explicitly depends on the gain. 
Consequently, the phase dynamics depend solely on the angular variables, 
effectively decoupling from both the amplitude and the carrier number. The 
resulting equation takes the form of the generalized time-delayed Kuramoto 
model~\cite{yeung:1999} with an additional phase shift $\tan^{-1}\alpha+\omega_0\tau$ and a dressed coupling enhanced by the amplitude–phase coupling $\alpha$: 
\begin{widetext}
\begin{align}\label{eq: time_delay_kuramoto}
    \dot{\Omega}_i(t) = \Delta_i - \sqrt{1+\alpha^2}\sum^{M}_{j=1}K_{ij}&\sin\big[\Omega_i(t)-\Omega_j(t-\tau)+\tan^{-1}\alpha+\omega_0\tau\big].
\end{align}
\end{widetext}

To derive the effective thermodynamic potential, we transform 
Eq.~\eqref{eq: time_delay_kuramoto} into the following gradient flow form:
$$\dot{\eta}_i(t)=-\frac{1}{\tau}\frac{d}{d\eta_i}U(\eta_i),$$
where the collective coupling influence of the other lasers on laser $i$ is 
contained in the thermodynamic potential $U(\eta_i)$. The time-delayed phase 
difference can be defined as
$$\eta_{ij}(t)=\Omega_i(t)-\Omega_j(t-\tau).$$
Associated with the steady-state synchronization solution, the phase variables
$\Omega_i(t)$ and $\Omega_j(t)$ are proximal to each other. The second-order 
tensor $\eta_{ij}(t)$ of the phase differences can then be approximated by a 
vector $\eta_i(t)$:
\begin{equation}\label{eq:eta_1}
     \eta_i(t) \approx \Omega_i(t) - \Omega_i(t-\tau).
\end{equation}
To proceed, we further assume a slow temporal evolution of the phase, so the 
finite difference can be approximated by the average derivative:
\begin{equation} \label{eq:eta_2}
    \frac{\eta_i(t)}{\tau}\approx\frac{\dot{\Omega}_i(t)+\dot{\Omega}_i(t-\tau)}{2}.
\end{equation}
Combining Eqs.~(\ref{eq:eta_1}) and (\ref{eq:eta_2}), we get
\begin{equation} \label{eq:eta_3}
    \eta_i(t)\approx \tau\dot{\Omega}_i(t)-\frac{1}{2}\tau\dot{\eta}_i(t).
\end{equation}
Substituting this relation into the generalized time-delayed Kuramoto model 
[Eq.~\eqref{eq: time_delay_kuramoto}], we obtain:
\begin{widetext}
\begin{align}
    &\dot{\eta}_i(t)=-\frac{1}{\tau}\bigg[2\eta_i(t)-2\tau\Delta_i+2\tau\sqrt{1+\alpha^2}\bigg(\sum^{M}_{j=1}K_{ij}\bigg)\sin[\eta_i(t)+\tan^{-1}\alpha+\omega_0\tau]\bigg].
\end{align}
The effective thermodynamic potential can then be identified as:
\begin{align}\label{eq:thermodynamic_potential}
     &U(\eta_i(t))= \eta^{2}_i(t)-2\tau\Delta_i\eta_i(t)-2\tau\sqrt{1+\alpha^2}k_{i}^{\textnormal{in}}\cos[\eta_i(t)+\tan^{-1}\alpha+\omega_0\tau],
\end{align}
\end{widetext}
and $k_{i}^{\textnormal{in}}\equiv \sum^{M}_{j=1}K_{ij}$ denotes the intrinsic coupling 
strength coming from other lasers to the $i$-th laser. This expression generalizes the result for 
a single laser~\cite{lenstra:1991}:
\begin{align}
U(\nu(t))=\nu^{2}(t)-2\tau\sqrt{1+\alpha^2}&\gamma\cos\big[\nu(t)+\tan^{-1}\alpha+\omega_0\tau\big],
\end{align}
where $\gamma$ is the self-coupling strength and 
$\nu(t)\equiv \Omega(t)-\Omega(t-\tau)$. 
The thermodynamic potential in Eq.~(\ref{eq:thermodynamic_potential}) indicates
that the effective coupling strength 
\begin{align} \nonumber
\mathbb{K}_{i}\equiv \tau\sqrt{1+\alpha^2}k_{i}^{\textnormal{in}} 
\end{align}
is governed by the time delay $\tau$, the amplitude-phase coupling factor $\alpha$, and the intrinsic coupling strength $k^{\textnormal{in}}_i$ 
(or $\gamma$ for a single laser). 

The local minima of the potential $U(\eta_i(t))$
can be determined through the first and second derivatives with respect to 
$\eta_{i}(t)$:
\begin{widetext} 
\begin{align}
    \frac{dU(\eta_i(t))}{d\eta_i(t)}&=\eta_i(t)-\tau\Delta_i+\mathbb{K}_i\sin\big[\eta_i(t)+\tan^{-1}\alpha+\omega_0\tau\big]=0,\label{eq: du_deta}\\
    \frac{d^2 U(\eta_i(t))}{d\eta^2_i(t)}&=1+\mathbb{K}_i\cos[\eta_i(t)+\tan^{-1}\alpha+\omega_0\tau]>0.\label{eq: d2u_deta2}
\end{align}
\end{widetext}
Setting $\omega_0\tau \equiv 2N\pi$, Eqs.~\eqref{eq: du_deta} and 
\eqref{eq: d2u_deta2} can be further simplified as
\begin{align}
    \eta_i - \mathbb{F}_i &= -\mathbb{K}_i\sin(\eta_i+\phi),\label{eq: ECM_criteria_1}\\
   & 1+\mathbb{K}_i\cos(\eta_i+\phi) > 0,\label{eq: ECM_criteria_2}
\end{align}
where $\mathbb{F}_i\equiv \tau\Delta_i$ denotes the effective detuning,
and the constant phase shift due to nonzero amplitude-phase coupling, is defined as 
$$\phi\equiv\textnormal{mod}(\tan^{-1}\alpha + \omega_0\tau,2\pi)\approx 0.4\pi.$$ 
The cross intersections from Eq.~\eqref{eq: ECM_criteria_1} correspond to the 
phase difference
$$\eta^{*}_{i}(t) \approx \Omega^{fit}_i\tau + \varphi_i(t)-\varphi_i(t-\tau)$$
for each laser $i$, where $\varphi(t)$ is the phase of the electric field after subtracting the fitted dynamical frequency component. In the case of complete frequency and phase synchronization among lasers, the solutions are expected to be independent of laser index $i$ and this simplifies to
$$\eta^{*}\approx \Omega^{fit}\tau=2\pi f_{\textnormal{final}}\tau,$$ 
under the approximation $\varphi(t)\approx \varphi(t-\tau)$, which renders $\eta$ time-independent. To compensate for frequency disorders and ensure that $\eta^{*}$ is independent 
of the laser index $i$, the effective coupling must be sufficiently strong: For an arbitrary laser $i$,
the effective coupling strength should exceed the effective detuning, i.e., $\mathbb{K}_i > |\mathbb{F}_i|$, or equivalently
\begin{align} \nonumber
k^{\textnormal{in}}_i > |\Delta_i|/\sqrt{1+\alpha^2}
\end{align}
so that all solutions $\eta_i$ fall within a similar and narrow range satisfying $1+\mathbb {K}_i\cos(\eta_i+\phi)>0$, and remain close to the initial mean frequency across general network configurations.

The variational principle stipulates that the system’s dynamics relax toward the global minimum or, when separated by appreciable barriers, to local minima of the potential. With or without $\alpha$, the steady-state solution structure remains qualitatively similar, but is shifted by the additional phase and modified by the dressed coupling due to amplitude-phase coupling $\alpha$. However, the LK equations determine the dynamical stability: nonzero amplitude–phase coupling $\alpha$ destabilizes the steady states and drives them into chaotic attractors, whereas in its absence the steady states remain stable, as shown in previous sections. 


%

\end{document}